\documentclass[aps,prb,reprint,superscriptaddress]{revtex4-1}
\usepackage[colorlinks=true,linkcolor=blue,citecolor=blue,urlcolor=blue]{hyperref}
\usepackage{amsmath}
\usepackage{amssymb} 
\usepackage[pdftex]{graphicx}
\usepackage{braket} 

\begin{document}

\newcommand{\RedText}[1]{\textcolor{red}{#1}}

\title{Cross-calibration of GaAs deformation potentials and gradient-elastic tensors using photoluminescence and nuclear magnetic resonance spectroscopy in GaAs/AlGaAs quantum dot structures.}


\author{E. A. Chekhovich}
\email[]{e.chekhovich@sheffield.ac.uk} \affiliation{Department of
Physics and Astronomy, University of Sheffield, Sheffield S3 7RH,
United Kingdom}
\author{I. M. Griffiths}
\affiliation{Department of Physics and Astronomy, University of
Sheffield, Sheffield S3 7RH, United Kingdom}
\author{M. S. Skolnick}
\affiliation{Department of Physics and Astronomy, University of Sheffield, Sheffield S3 7RH, United Kingdom}
\author{H. Huang}
\affiliation{Institute of Semiconductor and Solid State Physics,
Johannes Kepler University Linz, Altenbergerstr. 69, 4040 Linz,
Austria}
\author{S. F. Covre da Silva}
\affiliation{Institute of Semiconductor and Solid State Physics,
Johannes Kepler University Linz, Altenbergerstr. 69, 4040 Linz,
Austria}
\author{X. Yuan}
\affiliation{Institute of Semiconductor and Solid State Physics,
Johannes Kepler University Linz, Altenbergerstr. 69, 4040 Linz,
Austria}
\author{A. Rastelli}
\affiliation{Institute of Semiconductor and Solid State Physics,
Johannes Kepler University Linz, Altenbergerstr. 69, 4040 Linz,
Austria}

\date{\today}

\begin{abstract}
Lattice matched GaAs/AlGaAs epitaxial structures with quantum dots
are studied under static uniaxial stress applied either along the
$[001]$ or $[110]$ crystal directions. We conduct simultaneous
measurements of the spectral shifts in the photoluminescence of
the bulk GaAs substrate, which relate to strain via deformation
potentials $a$ and $b$, and the quadrupolar shifts in the
optically detected nuclear magnetic resonance spectra of the
quantum dots, which relate to the same strain via the
gradient-elastic tensor $S_{ijkl}$. Measurements in two uniaxial
stress configurations are used to derive the ratio
$b/a=0.241\pm0.008$ in good agreement with previous studies on
GaAs. Based on the previously estimated value of $a\approx-8.8$~eV
we derive the product of the nuclear quadrupolar moment $Q$ and
the $S$-tensor diagonal component in GaAs to be
$QS_{11}\approx+0.76\times10^{-6}$~V for $^{75}$As and
$QS_{11}\approx-0.37\times10^{-6}$~V for $^{69}$Ga nuclei. In our
experiments the signs of $S_{11}$ are directly measurable, which
was not possible in the earlier nuclear acoustic resonance
studies. Our $QS_{11}$ values are a factor of $\sim$1.4 smaller
than those derived from the nuclear acoustic resonance experiments
[Phys. Rev. B 10, 4244 (1974)]. The gradient-elastic tensor values
measured in this work can be applied in structural analysis of
strained III-V semiconductor nanostructures via accurate modelling
of their magnetic resonance spectra.
\end{abstract}


\maketitle

\section{Introduction}

Electronic and optical properties of semiconductors depend
strongly on the symmetry of the underlying crystal structure
\cite{ChuangBook,SunBook}. Many technologically important
semiconductors, such as Si, Ge, GaAs, InP have high crystal
symmetry belonging to the cubic crystal system. Elastic
deformation (strain) induced by external stress or internal
morphology leads to reduction of the crystal symmetry, resulting
in significant modification of the optical and electronic
properties. Strain-induced effects not only serve as a tool in
studying the physics and structure of semiconductors, but have
already found several important applications, including pressure
sensors and transducers, as well as MOSFET transistors and
semiconductor lasers with improved performance. Semiconductor
technologies under development also involve strain effects. One
example is quantum information technologies based on semiconductor
quantum dots, where strain is used both in self assembly growth of
the quantum dot nanostructures and for tuning their properties
\cite{PhysRevLett.104.067405,doi:10.1063/1.1566463,doi:10.1063/1.2204843,doi:10.1002/adma.201200537,doi:10.1002/pssb.201100775}.

The changes in semiconductor electronic properties induced by
strain originate from the changes in orientations and overlaps of
the electronic orbitals. One manifestation of these changes is in
the shifts of the energies of the electronic bands, and in lifting
of their degeneracies. In GaAs the strain-induced modification of
the electronic structure can be described by four parameters --
the deformation potentials $a_c$ and $a_v$ describe the overall
shift of the conduction and valence bands respectively, while $b$
and $d$ describe lifting of degeneracy and splitting in the
valence band. Deformation potentials of GaAs have been
measured\cite{PhysRevLett.16.942,PhysRev.161.695,BENDORIUS19701111,PhysRevB.15.2127,QIANG19901087,MAIR1998277}
using photoluminescence, photoreflectance, and electroreflectance
techniques. The most consistent experimental and
theoretical\cite{PhysRevB.37.8519,PhysRevB.39.1871,PhysRevB.60.5404,PhysRevB.84.035203,0022-3727-49-8-085104}
results are available for the combination $a=a_c+a_v$, which
describes the change of the direct band gap in a deformed crystal
\cite{Vurgaftman2001,Adachi2009}. The largest uncertainty is
associated with the individual values of $a_c$ and $a_v$. The $b$
and $d$ have been measured as well, although the values quoted in
different reports vary by as large as a factor of $\sim$2.

The same strain-induced changes of the electronic bonds are
responsible for non-zero electric field gradients (EFGs) at the
sites of the atomic nuclei (EFGs vanish in an unstrained crystal
with cubic symmetry). This effect can be observed as quadrupolar
splitting of the nuclear magnetic resonance (NMR) spectra of the
nuclei with spin $I>1/2$. The relation between strain and EFG is
described by a fourth rank ''gradient-elastic'' tensor $S_{ijkl}$,
which can be parameterized by two components $S_{11}$ and $S_{44}$
in case of cubic crystal symmetry. The need for accurate
$S_{ijkl}$ values have reemerged recently in view of using NMR for
non-destructive structural analysis of nanoscale semiconductor
structures
\cite{chekhovich2012,MunschNMR,PhysRevB.82.081308,PhysRevB.93.045301,PhysRevB.89.125304,PhysRevB.85.115313,PhysRevB.90.205425}
as well as exploring the effect of nuclear quadrupolar interaction
on coherent electron-nuclear spin dynamics in solid state
qubits\cite{chekhovich2015,wust2016,botzem2016}.

The initial measurements of $S_{ijkl}$ in various crystal
materials used static straining, but their accuracy suffered since
quadrupolar spectral shifts were not resolved and could only be
observed as broadening of the NMR spectra \cite{PhysRev.107.953}.
In later experiments more reliable measurements were achieved as
NMR spectra with resolved quadrupolar satellites could be obtained
under static strain \cite{PhysRev.150.546,Bogdanov1967}, but in
the particular case of GaAs, no accurate estimates of $S_{ijkl}$
could be derived \cite{Bogdanov1968}. Sundfors et. al. have
derived $S_{ijkl}$ for a wide range of
materials\cite{PhysRev.177.1221,PhysRevB.10.4244,PhysRevB.12.790,PhysRevB.13.4504}
including GaAs and other III-V semiconductors. The experiments in
these studies relied on measuring absorbtion of the acoustic waves
rather than direct detection of the quadrupolar shifts in NMR
spectra. In a more recent study optically detected NMR was
measured in a GaAs/AlGaAs quantum well under static bending strain
\cite{Guerrier1997}. Quadrupolar shifts were resolved for
$^{75}$As and were found to be consistent with the results of
acoustic resonance
measurement\cite{PhysRev.177.1221,PhysRevB.10.4244}. However, the
induced deformation was comparable to the built-in strain, the
accuracy of strain measurement was limited, and oblique magnetic
field configuration meant that individual $S_{ijkl}$ components
were not derived explicitly.

Here we study GaAs/AlGaAs quantum dot (QD) structures and perform
simultaneous measurements of optically detected NMR on individual
QDs and photoluminescence of free excitons in bulk GaAs substrate
in a sub-micrometer vicinity of the QD. Large elastic deformations
exceeding built-in strains by more than an order of magnitude are
induced by stressing the samples mechanically. Optically detected
NMR reveals spectra with well-resolved quadrupolar satellites, so
that quadrupolar shifts are measured with an accuracy of $\pm1\%$.
Using the commonly accepted value for deformation potential $a$,
the energy shifts in the free exciton photoluminescence of the
GaAs substrate are used to measure the magnitude of the same
strain field that is probed via QD NMR. From these dual
measurements we are able to relate elastic strain to the directly
measured nuclear spin quadrupolar shifts and deduce the $S_{11}$
components of the gradient-elastic tensor of $^{75}$As and
$^{69}$Ga in GaAs. Our accurate measurements reveal $S_{11}$ that
are $\sim$30\% smaller than the only direct measurement based on
nuclear acoustic resonance \cite{PhysRevB.10.4244}. The $S_{11}$
constants derived in this work can be used directly in analysing
and predicting the nuclear quadrupolar effects in GaAs-based
semiconductor nanostructures. Furthermore, since gradient-elastic
tensors describe modification of the electronic orbitals in the
vicinity of the nucleus, the accurate experimental $S_{11}$ values
can be used as a reference in fitting the calculated parameters in
electronic band-structure modelling.

\section{Strain effects in $GaAs$: definitions}
The electronic band structure of a bulk crystal can be described
by the Luttinger model where the effects of strain are taken into
account by the Bir-Pikus Hamiltonian \cite{SunBook,ChuangBook}.
The optical recombination properties of GaAs are determined mainly
by the states with momentum $k\approx0$ corresponding to the
centre of the Brilluoin zone which simplifies the analysis. The
bottom of the conduction band is two-fold degenerate due to the
electron spin, and as such remains degenerate under strain. The
only effect of strain on the conduction band is an overall energy
shift $a_c\epsilon_h$, which depends only on the hydrostatic part
of the strain tensor
$\epsilon_h=\epsilon_{xx}+\epsilon_{yy}+\epsilon_{zz}$ (here, and
throughout the text we use coordinate frame aligned with the cubic
crystal axes $x\parallel[100]$, $y\parallel[010]$,
$z\parallel[001]$). In case of GaAs $a_c<0$, so that under
compressive strain ($\epsilon_h<0$) the conduction band energy
increases.

Without strain, the cubic symmetry of GaAs results in a four-fold
degeneracy at the top of the valence band. At small strains the
energies of the valence band at $k=0$ can be adequately described
without coupling to the split-off band, which reduces the model to
a 4$\times$4 Hamiltonian with a straightforward analytical
solution. Strain does not break time reversal symmetry, and thus
at most can split the valence band into two states each with a
two-fold degeneracy. The valence band energy shifts are
$-a_v\epsilon_h\pm\sqrt{b^2\epsilon_b^2+\frac{3}{4}b^2\epsilon_{\eta}^2+d^2\epsilon_s^2}$,
where $\epsilon_{b}=\epsilon_{zz}-(\epsilon_{xx}+\epsilon_{yy})/2$
is the ''biaxial'' component of the shear strain, and we denote
$\epsilon_{\eta}=\epsilon_{xx}-\epsilon_{yy}$ and
$\epsilon_{s}^2=\epsilon_{xy}^2+\epsilon_{yz}^2+\epsilon_{xz}^2$.
It is commonly accepted that under compressive hydrostatic strain
($\epsilon_h<0$) the valence band moves to lower energy,
corresponding to $a_v<0$ with the sign convention used
here\cite{Vurgaftman2001}. The energy of the photoluminescence
photons (measurable experimentally) is the difference of the
conduction and valence band energies and can be written as
\begin{equation}
\begin{split}
E_\textrm{PL}=E_g+a\epsilon_h\pm\sqrt{b^2\epsilon_b^2+\frac{3}{4}b^2\epsilon_{\eta}^2+d^2\epsilon_s^2},
\end{split}
\label{Eq:EPL}
\end{equation}
where $E_g$ is the direct bandgap energy of unstrained GaAs. Under
uniaxial compressive strain along $z$ (characterized by
$\epsilon_{zz}<0$ and $\epsilon_{xx}=\epsilon_{yy}>0$) the
transition with lower PL energy corresponds to the valence band
light holes (LH) with momentum $j_z=\pm1/2$, while higher PL
energy corresponds to the heavy holes with momentum $j_z=\pm3/2$.

In any crystal in equilibrium the electric field at the atomic
nucleus site is zero. However the gradients of the electric field
components are not necessarily zero and are described by a
symmetric second rank tensor $V_{ij}$ of the second spatial
derivatives of the electrostatic potential $V$. In a crystal with
cubic symmetry $V_{ij}$ vanishes at the nuclear sites, but when
the crystal is strained, electric field gradients arise and in
linear approximation are related to the strain tensor
$\epsilon_{kl}$ via $V_{ij}=S_{ijkl}\epsilon_{kl}$. A nucleus with
a non-zero electric quadrupolar moment $Q$ interacts with the
electric field gradients. In a simplest case of high static
magnetic field the effect of the quadrupolar interaction is to
split the NMR transition into a multiplet of transitions between
the states whose spin projections onto magnetic field differ by
$\pm1$. In case of spin $I=3/2$ nuclei and static magnetic field
directed along the $z$ axis a triplet of equidistant NMR
frequencies is observed with splitting
\cite{Volkoff1952,PhysRev.107.953,Guerrier1997}:
\begin{equation}
\begin{split}
\nu_{Q}=\frac{eQ}{2 h}S_{11}\epsilon_{b},\\
\end{split}
\label{Eq:nuQ}
\end{equation}
where $e>0$ is the elementary charge, $h$ is the Planck constant
and we used Voigt notation for the component of the gradient
elastic tensor $S_{11}=S_{xxxx}=S_{yyyy}=S_{zzzz}$. (For a
detailed derivation see Appendix~\ref{Appendix:QNuc}). Unlike the
free exciton energies measured in PL spectroscopy
(Eq.~\ref{Eq:EPL}), the shifts measured in NMR spectra
(Eq.~\ref{Eq:nuQ}) are not sensitive to the hydrostatic strain and
depend only on shear strains of a particular symmetry (described
by $\epsilon_{b}$). This property is exploited in this work to
cross-calibrate the magnitudes of $S_{11}$ and deformation
potentials.

\section{Samples and experimental techniques}

\begin{figure}[h]
\includegraphics[viewport=1 1 562 511, clip=true, width=1.0\linewidth]{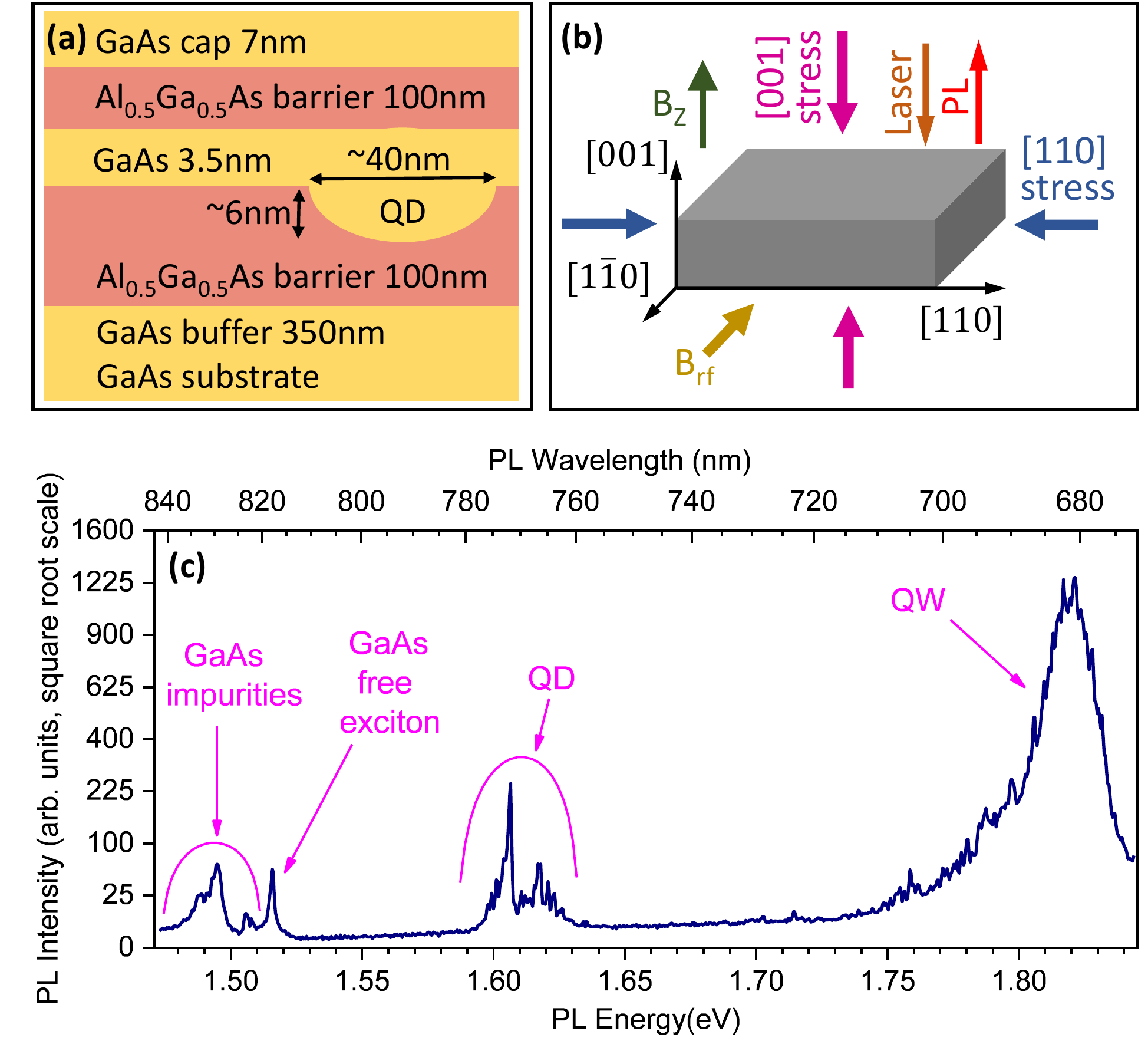}
\caption{\label{Fig:FigSamples} (a) Sample structure showing the
sequence of GaAs and Al$_{0.5}$Ga$_{0.5}$As epitaxial layers. (b)
Schematic of the experiment geometry showing orientation of a
sample, direction of the static magnetic field $B_z$, radio
frequency field $B_{rf}$, and the direction of the
photoluminescence excitation and collection. External stress is
applied either along the $[001]$ direction or the $[110]$
direction. (c) Typical photoluminescence spectrum at $B_z=0$
showing emission from the quantum well (QW), a single quantum dot
(QD), as well as emission from the GaAs substrate which includes
free exciton emission and impurity-induced recombination.
Square-root vertical scale is used to reveal weak spectral
features.}
\end{figure}

The structure studied in this work was grown using molecular beam
epitaxy. The schematic cross section is shown in
Fig.~\ref{Fig:FigSamples}(a). The first step in the growth is the
deposition of a 350~nm thick buffer GaAs layer onto an undoped
$\sim$0.35~mm thick $(001)$-oriented GaAs wafer. This is followed
by the growth of a 100~nm thick bottom barrier
Al$_{0.5}$Ga$_{0.5}$As layer. Aluminium droplets are then grown
and used to etch nanoholes in the bottom barrier
\cite{Atkinson2012}. A typical nanohole is $\sim$40~nm in diameter
and $\sim$5~nm deep. A layer of GaAs with a nominal thickness of
3.5~nm is then deposited, resulting in formation of quantum dots
(QDs) due to filling up of the nanoholes, as well as formation of
a quantum well (QW) layer. A 100~nm thick top
Al$_{0.5}$Ga$_{0.5}$As barrier layer is then grown, followed by a
7~nm thick cap layer.

The structure was cleaved into small parallelepiped pieces with
dimensions of $\sim0.9\times1.5\times0.35$~mm along the $[110]$,
$[1\bar{1}0]$ and $[001]$ directions respectively. Three samples
were prepared. The first sample was as grown (unstressed). The
second sample was glued between two flat titanium surfaces and
stressed compressively along the $[110]$ direction using titanium
screw and nut that press the two titanium surfaces towards each
other. The third sample was glued between the bottom titanium flat
surface and the top sapphire flat surface to be stressed
compressively along the $[001]$ growth direction. All of the
samples were studied in a configuration shown in
Fig.~\ref{Fig:FigSamples}(b). Magnetic field up to 10~T was
aligned along the $z$-axis ($[001]$) within $\pm2^{\circ}$, which
is also the direction of the laser excitation and
photoluminescence (PL) collection. For the sample stressed along
the $[001]$ direction, optical excitation and PL propogated
through the sapphire glass.

All experiments are conducted in a helium bath cryostat at a
sample temperature $\sim$4.2~K. A small copper coil is mounted
close to the sample and is used to generate radiofrequency
magnetic field $B_{rf}$ along the $[1\bar{1}0]$ direction in the
NMR experiments. Quantum dot NMR spectra are measured using
optical hyperpolarization of the nuclear spins (via circularly
polarized laser excitation) and optical detection of the electron
hyperfine shifts. The signals of the quadrupolar nuclei are
enhanced using ''inverse'' NMR technique \cite{chekhovich2012}. A
detailed description and analysis of the relevant NMR methods has
been reported previously \cite{chekhovich2012}, and is not
repeated here: in this work we use these techniques as a tool that
gives an accurate spectral distribution of the resonant
frequencies of the nuclei within the volume of an individual
quantum dot. The excitation laser is focused into a spot of
$\sim$1~$\mu$m in diameter, so that carriers are generated
simultaneously in the QW, the GaAs buffer layer, and the QDs
within the area of the laser spot. The photoluminescence signal is
collected and analyzed with a grating spectrometer and a charge
coupled device (CCD) camera.

A typical broadband PL spectrum measured under HeNe laser
excitation ($632.8$~nm) is shown in Fig.~\ref{Fig:FigSamples}(c).
Spectral features observed include emission from the QW
($\sim$1.85~eV), free exciton emission of the bulk GaAs buffer and
substrate layers ($\sim$1.515~eV), impurity-induced PL of bulk
GaAs ($\sim$1.48-1.51~eV) including bound excitons as well as
recombination involving donor and acceptor states
\cite{PhysRevB.58.R13403,PhysRev.184.811,doi:10.1063/1.336559,doi:10.1063/1.372104,PhysRevB.73.035205,PhysRevB.7.4568}.
Quantum dot emission is observed at $\sim$1.60-1.63~eV and
consists of several narrow spectral lines corresponding to
different exciton states of a single QD. Since photoluminescence
is excited only in a small area of the sample, the spectrum of
GaAs free exciton can be used to probe local strain fields in a
$\sim$1~$\mu$m sized spot. Moreover, NMR is detected from the
spectral shifts in the QD emission and thus samples an even
smaller nanometer-sized part of the optically excited area. In
this way it is ensured that GaAs PL spectroscopy and QD NMR sample
the same strain field.

\section{Experimental results and analysis}

\subsection{Effect of strain on GaAs photoluminescence and nuclear magnetic resonance spectra\label{SubSec:RawData}}

\begin{figure*}[t]
\includegraphics[width=1.0\linewidth]{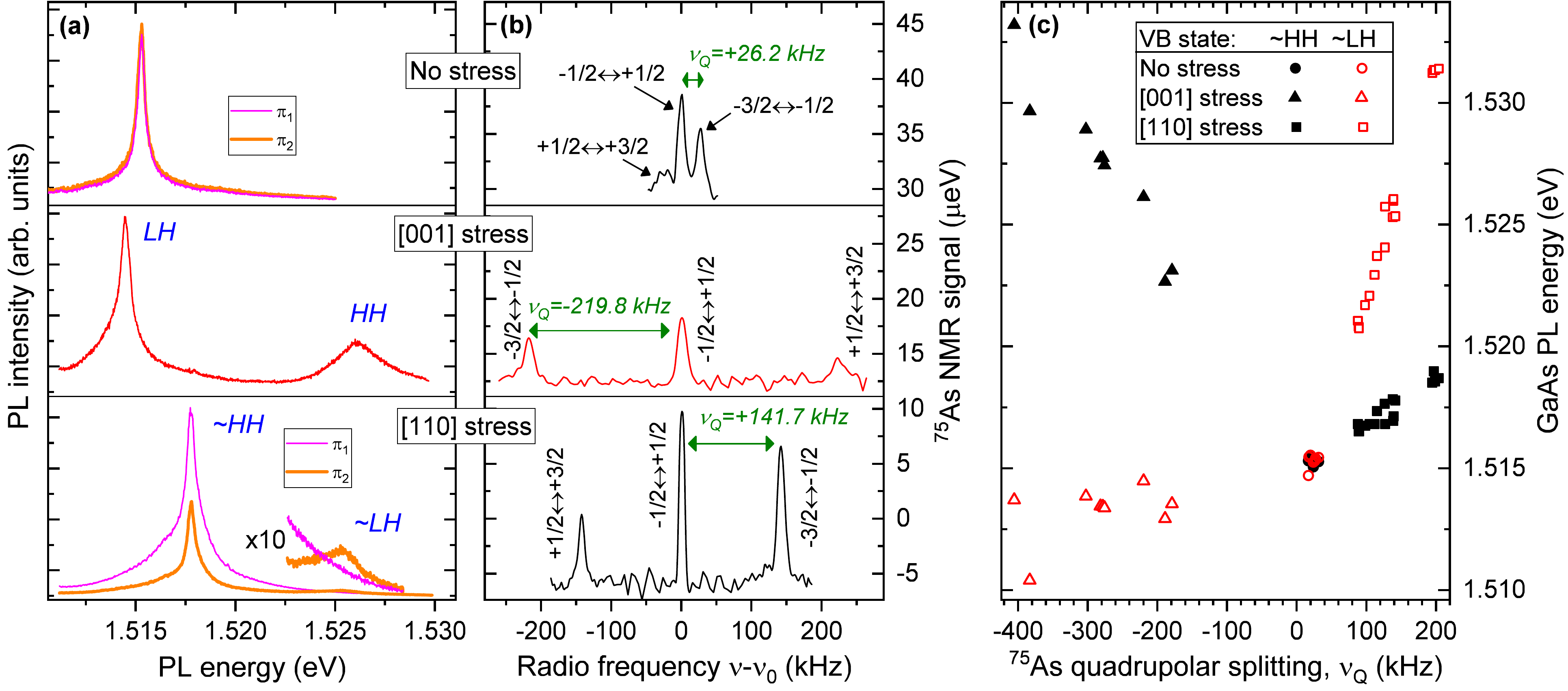}
\caption{\label{Fig:FigRawData} Effect of strain on bulk GaAs
photoluminescence (PL) and quantum dot NMR spectra. (a)
Free-exciton PL from a GaAs substrate measured at $B=$0~T under
excitation with a photon energy $\sim$1.54~eV and intensity
$\sim$5$\times10^6$~W/m$^2$ in an unstressed sample (top), sample
stressed along $[001]$ (middle), and sample stressed along $[110]$
(bottom). In an unstressed sample emission has negligible
splitting between peaks detected in two orthogonal linear
polarizations ($\pi_1$, $\pi_2$). Stress along $[001]$ splits
luminescence spectrum into two non-polarized peaks corresponding
to emission of light (LH) and heavy (HH) hole excitons. Stress
along $[110]$ splits luminescence spectrum into a stronger peak
with partial linear polarization corresponding to emission from a
predominantly HH exciton, and a weak linearly polarized peak from
a predominantly LH exciton (parts of the spectra with $\times10$
vertical magnification are shown to reveal the $\sim$LH peak). (b)
Nuclear magnetic resonance spectra of $^{75}$As nuclei measured
with $\sigma^+$ polarized optical excitation at $B_z=$8~T
($\nu_0\approx$58.46~MHz) on GaAs/AlGaAs quantum dots from the
same spots as GaAs luminescence spectra in (a). Well-resolved NMR
triplets arising from quadrupolar effects are observed. In an
unstressed sample small quadrupolar shift $\nu_Q$ is observed due
to the residual strain of the GaAs/AlGaAs heterostructure. Under
$[001]$ ($[110]$) stress, the resulting strain shifts the
$-3/2\leftrightarrow-1/2$ satellite transition to lower (higher)
frequency corresponding to negative (positive) $\nu_Q$. (c)
Energies of predominantly HH (solid symbols) and LH (open symbols)
bulk GaAs PL peaks plotted against quadrupolar shift $\nu_Q$
measured in multiple quantum dots in an unstressed (circles),
$[001]$-stressed (triangles), and $[110]$-stressed (squares)
samples. GaAs PL energies and NMR frequencies are derived from the
spectra using Lorentzian and Gaussian peak fitting respectively.
Variation of $\nu_Q$ and PL energies in each sample is due to the
inhomogeneity of strain across the sample surface -- the exception
is the four points at $\nu_Q\approx+200$~kHz that were measured at
an increased stress along $[110]$.}
\end{figure*}

Figure~\ref{Fig:FigRawData}(a) shows GaAs free exciton PL spectra
measured in three different samples at $B_z$=0, while
Fig.~\ref{Fig:FigRawData}(b) shows $^{75}$As NMR spectra measured
at $B_z$=8~T from the QDs in the same optically excited spots as
in (a). Since the size of the optically excited spot is much
smaller than the size of the sample, and the stiffness tensors of
GaAs and AlAs are very similar \cite{Vurgaftman2001} all
significant variations of strain induced by external stress occur
on length scales that are much larger than the studied spot size.
As a result the two types of spectroscopy probe the same strain
field.

Bulk GaAs PL is measured with laser excitation intensity
$\sim$5$\times10^6$~W/m$^2$. On the one hand it is high enough to
saturate the impurity-induced PL and make free exciton emission
dominant, while on the other hand it is low enough to avoid
excessive spectral broadening. In an unstressed sample PL is
detected with a variable orientation of linear polarization: the
top two spectra in Fig.~\ref{Fig:FigRawData}(a) are measured along
the orthogonal polarization axes and reveal very small
polarization degree and a negligible splitting. This is expected
for unstrained GaAs PL, since the valence band state at $k=0$ is
four-fold degenerate. The corresponding NMR spectrum
(Fig.~\ref{Fig:FigRawData}(b), top) reveals a triplet of lines
with a small quadrupolar splitting $|\nu_Q|\approx$26.2~kHz, most
likely related to the strain arising from the residual lattice
mismatch of the GaAs and Al$_{0.5}$Ga$_{0.5}$As layers. Each line
of the triplet corresponds to an individual dipolar nuclear spin
transition $I_z\leftrightarrow I_{z+1}$ as labeled in
Fig.~\ref{Fig:FigRawData}(b).

For the sample stressed along $[001]$, GaAs free exciton PL is
split into two non-polarized lines (Fig.~\ref{Fig:FigRawData}(a),
middle). This is expected, since deformation along $[001]$ lifts
the degeneracy and splits the state at the top of the valence band
into a two-fold degenerate state with momentum projection
$j=\pm1/2$ corresponding to the light holes (LH), and a two-fold
degenerate state with $j=\pm3/2$ corresponding to the heavy holes
(HH). The effect of strain is also manifested in NMR through a
significantly larger triplet splitting $|\nu_Q|\approx$219.8~kHz
(Fig.~\ref{Fig:FigRawData}(b), middle).

The stress along $[110]$ also splits the four-fold degenerate top
of the valence band into two doublets. These however are no longer
pure heavy and light hole states, and their recombination results
in a linearly polarized PL (Fig.~\ref{Fig:FigRawData}(a), bottom).
The peak at $\sim$1.518~eV is partially linearly polarized and
corresponds to the state with predominantly heavy hole character
($\sim$HH). By contrast, the peak at $\sim$1.525~eV is strongly
polarized and corresponds to a predominantly light hole state
($\sim$LH). The intensity of the $\sim$LH peak is reduced due to
the relaxation into the $\sim$HH state. The NMR triplet splitting
(Fig.~\ref{Fig:FigRawData}(b), bottom) is also significantly
larger than in an unstressed sample with
$|\nu_Q|\approx$141.7~kHz.

The measurements of GaAs free exciton PL and $^{75}$As NMR were
repeated on multiple spots in all three samples and spectra
similar to those shown in Fig.~\ref{Fig:FigRawData}(a,b) were
observed. For each spot PL energies and NMR frequencies were
derived by fitting the spectral peaks. The resulting summary in
Fig.~\ref{Fig:FigRawData}(c) shows PL energies of $\sim$HH/HH
(solid symbols) and $\sim$LH/LH (open symbols) excitons as a
function of the quadrupolar splitting $\nu_Q$ in an unstressed
(circles), $[001]$-stressed (triangles), and $[110]$-stressed
(squares) samples. It can be seen that in the unstressed sample
$\nu_Q$ varies in a small range between 15 and 30~kHz, due to the
differences in the residual strains in the individual quantum
dots, while GaAs PL peak energy varies in a small range between
1.5145 and 1.5155~eV, most likely due to the local residual
strains arising from crystal imperfections. The spectral shifts in
the stressed samples are significantly larger than the random
variations in the unstressed sample. There is a clear trend in
Fig.~\ref{Fig:FigRawData}(c) that larger quadrupolar shifts are
correlated with larger GaAs PL energy shifts. On the other hand,
the stress-induced spectral shifts (both in PL and NMR) vary
across the surface area of the sample, since non-uniform contact
between the sample and the titanium stress mount leads to spatial
non-uniformity of the stress and strain fields. However, these
non-uniformities have characteristic lengths much larger than the
laser excitation spot, so that the strain detected in optical PL
and NMR spectra can be treated as constant for each individual
spot.

\begin{figure}[t]
\includegraphics[width=1.0\linewidth]{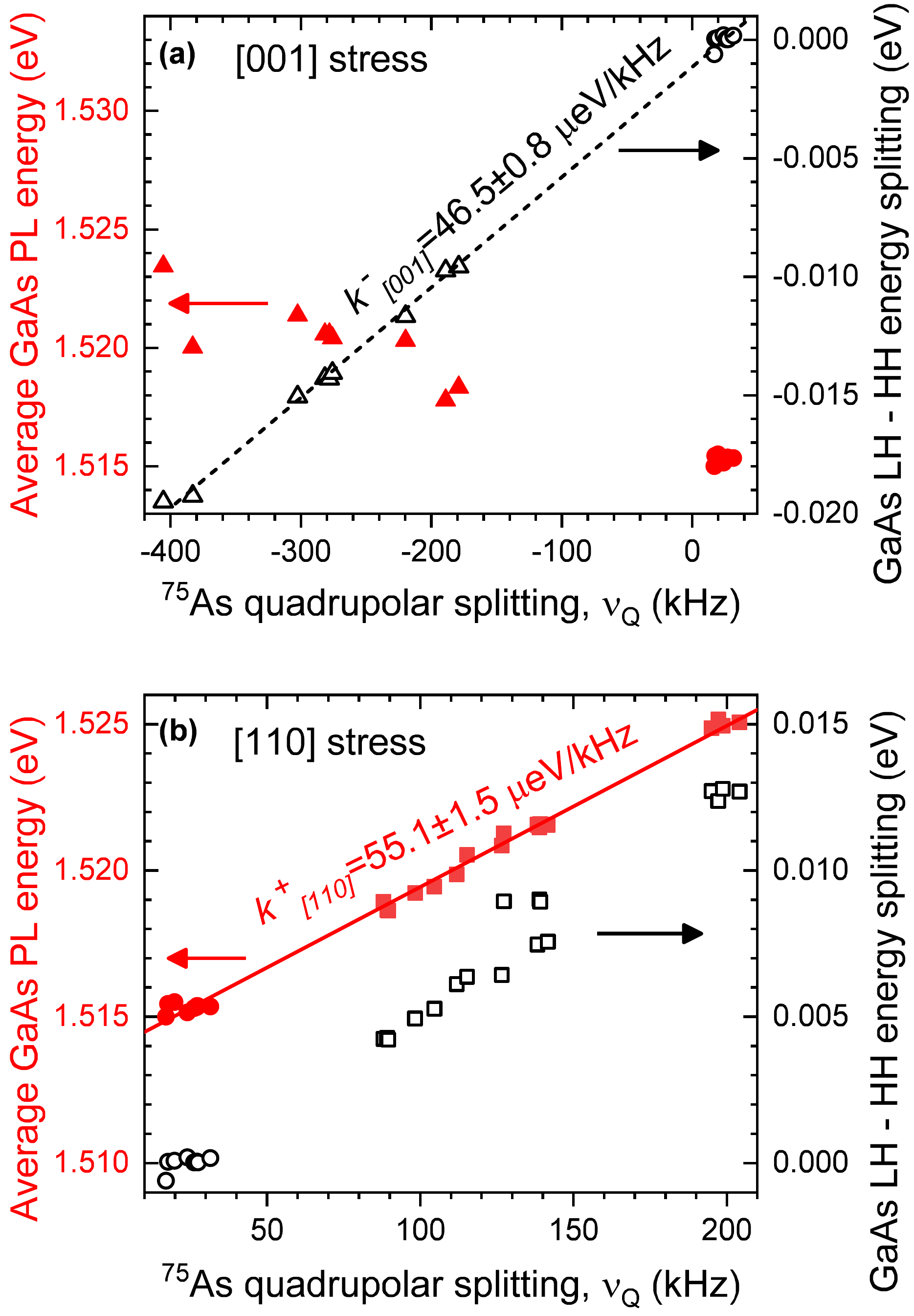}
\caption{\label{Fig:FigEPLMeanDiff} Data of
Fig.~\ref{Fig:FigRawData}(c) plotted in terms of the arithmetic
average PL energy of the LH and HH excitons (solid symbols, left
scales) and the difference of the LH and HH exciton PL energies
(open symbols, right scales) as a function of $^{75}$As
quadrupolar shift $\nu_Q$. (a) Results for the unstressed
(circles) and $[001]$-stressed (triangles) samples. LH-HH
splitting is well described by a linear function with a slope
$k_{[001]}^-=46.5\pm0.8~\mu$eV/kHz (dashed line). (b) Results for
the unstressed (circles) and $[110]$-stressed (squares) samples.
The average LH-HH PL energy is well described by a linear function
with a slope $k_{[110]}^+=55.1\pm1.5~\mu$eV/kHz (solid line).}
\end{figure}

For the purpose of quantitative analysis it is convenient to
re-plot the data of Fig.~\ref{Fig:FigRawData}(c) in a different
form. This is shown in Fig.~\ref{Fig:FigEPLMeanDiff} where the
average energy of LH and HH (solid symbols, left scales) as well
as the splitting of the LH and HH (open symbols, right scales) are
plotted as a function of $\nu_Q$ for the $[001]$-stressed (a) and
$[110]$-stressed (b) samples.

In case of the $[001]$-stressed sample
[Fig.~\ref{Fig:FigEPLMeanDiff}(a)], the average energy of LH and
HH shows significant random variations. By contrast, the LH-HH
splitting is very well described by a linear dependence on
$\nu_Q$. The best fit is shown by a dashed line in
Fig.~\ref{Fig:FigEPLMeanDiff}(a) and the slope is
$k_{[001]}^-=46.5\pm0.8 \mu$eV/kHz (95\% confidence interval). The
situation is reversed for the sample stressed along $[110]$ as
shown in Fig.~\ref{Fig:FigEPLMeanDiff}(b). While the LH-HH
splitting shows variations, the dependence of the average LH and
HH recombination energies is well described by a linear function
(solid line) with a fitted slope $k_{[110]}^+=55.1\pm1.5
\mu$eV/kHz. As we show below, such a difference between the cases
of $[001]$-stressed and $[110]$-stressed samples is not a
coincidence. With some basic assumptions about the spatial
distribution of strain in the stressed samples the measured
$k_{[001]}^-$ and $k_{[110]}^+$ values are used to derive the
gradient elastic tensor component $S_{11}$ as show in
Sec.~\ref{Subsec:SDerivation}. Prior to this derivation, in the
next subsections we present analysis of the properties of the
gradient-elastic tensor that require no assumptions about strain
configuration.

It is worth noting that rigorous analysis of bulk GaAs PL spectra
requires taking into account electron-hole exchange interaction
and polariton effects. However, these effects are of the order of
$\sim$0.25~meV, which is significantly smaller than the strain
induced spectral shifts observed here. More importantly, it has
been shown that the strain-induced spectral shifts of all the PL
components are well described by the free electron and hole
deformation potentials \cite{PhysRevB.7.4568}. Since our
subsequent analysis relies only on the ratios of the
strain-induced PL and NMR spectral shifts (rather than absolute
GaAs PL energies), it is sufficient to use a simplified
''free-exciton'' description of the GaAs PL ignoring polariton
effects.

\subsection{Measurement of the sign of the $S$-tensor components\label{Subsec:SSign}}

\begin{figure}[t]
\includegraphics[width=1.0\linewidth]{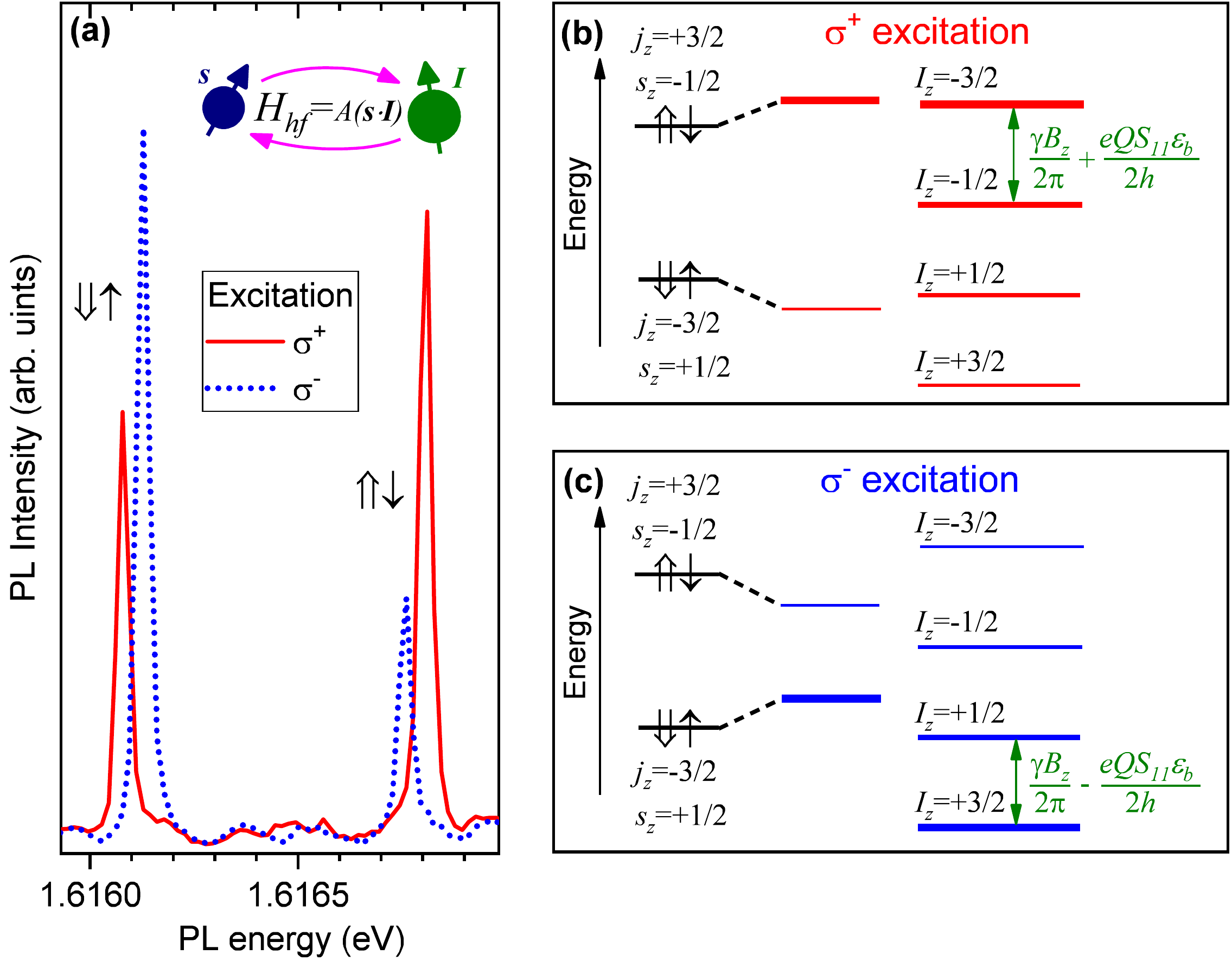}
\caption{\label{Fig:FigSSign} Derivation of the sign of the
gradient elastic tensor. (a) Photoluminescence spectra of a QD
neutral exciton measured at $B_z$=8~T under $\sigma^+$ (solid
line) and $\sigma^-$ (dashed line) polarized excitation at
1.65~eV. The $\sigma^+$ excitation predominantly populates the
exciton state with hole spin $s_z=+3/2$ ($\Uparrow$) and electron
spin $s_z=-1/2$ ($\downarrow$), while $\sigma^-$ excitation
predominantly populates the $\Downarrow\uparrow$ exciton. Electron
spin $\mathbf{s}$ can be transferred to a nuclear spin
$\mathbf{I}$ via hyperfine interaction (inset) resulting in
dynamic nuclear polarization, which in turn leads to hyperfine
shifts of the exciton transitions. (b) Schematic of spin effects
under $\sigma^+$ excitation which increases population of the
$s_z=-1/2$ excitons (thicker horizontal line) and reduces
population of the $s_z=+1/2$ exciton (thinner line). Dynamic
nuclear polarization enhances the population of the $I_z=-3/2$
nuclear spin states. This is observed as a hyperfine shift of the
$s_z=-1/2$ excitons to higher energy and enhanced amplitude of the
$-3/2\leftrightarrow-1/2$ NMR transition with frequency
$\frac{\gamma B_z}{2 \pi}+\frac{eQ}{2 h}S_{11}\epsilon_{b}$. (c)
Excitation with $\sigma^-$ light leads to the opposite sign of
electron and nuclear spin polarizations, enhancing the
$+1/2\leftrightarrow+3/2$ NMR transition at frequency
$\frac{\gamma B_z}{2 \pi}-\frac{eQ}{2 h}S_{11}\epsilon_{b}$. By
matching the signs of the exciton spectral shifts in (a) and the
sign of the NMR spectral shifts [cf. Fig.~\ref{Fig:FigRawData}(b)]
it is possible to deduce the sign of the gradient elastic tensor
component $S_{11}$ (see Sec.~\ref{Subsec:SSign}).}
\end{figure}

We now show how the sign of the gradient-elastic tensor can be
determined directly, if it is possible to identify spin
projections of the nuclear spin states corresponding to each NMR
transitions. The nuclear spin states can be identified from the
hyperfine interaction effects, if the sing of the electron spin
polarization is known. In order to define the sign of the electron
spin polarization we start by considering the signs of the carrier
$g$-factors. The electron $g$-factor in the studied QDs
\cite{PhysRevB.93.165306} as well as in thin GaAs/AlGaAs quantum
wells \cite{PhysRevB.45.3922} is small and the shifts of the
excitonic levels induced by magnetic field along the growth axis
are dominated by the hole Zeeman effect. The sign of the hole
$g$-factor \cite{PhysRevB.45.3922,PhysRevB.93.165306} is such that
at positive magnetic field $B_z>0$ the exciton with a positive
(negative) hole momentum projection $j_z$=$+3/2$ ($-3/2$) labeled
$\Uparrow$ ($\Downarrow$) has higher (lower) energy. In order to
be optically active the high- (low-) energy exciton must have
electron spin projection $s_z$=$-1/2$ ($+1/2$) denoted
$\downarrow$ ($\uparrow$). Figure~\ref{Fig:FigSSign}(a) shows PL
spectra of a neutral exciton in a typical QD at $B_z=8$~T measured
under $\sigma^+$ and $\sigma^-$ optical excitation at
$\sim$1.65~eV. Each PL spectrum is a doublet of optically allowed
(''bright'') excitons, with high- (low-) energy Zeeman component
corresponding to recombination of a $\Uparrow\downarrow$
($\Downarrow\uparrow$) exciton.

Two effects are observed under circularly polarized excitation in
Fig.~\ref{Fig:FigSSign}(a): (i) the emission intensity of the
high-(low-) energy Zeeman component is enhanced under $\sigma^+$
($\sigma^-$) excitation, and (ii) the Zeeman splitting increases
(decreases) under $\sigma^+$ ($\sigma^-$) excitation due to the
buildup of nuclear spin polarization. These two effects are
related: to understand their origin we first consider the case of
$\sigma^+$ excitation [Fig.~\ref{Fig:FigSSign}(b)], which
generates predominantly $\Uparrow\downarrow$ excitons. During
repeated optical excitation the $s_z=-1/2$ electrons transfer
their polarization to the nuclei of the dot via the flip-flop
process \cite{chekhovich2017} enabled by the hyperfine
interaction. Since the flip-flops are spin-conserving, the nuclei
become predominantly polarized into the states with negative spin
$I_z<0$. The net nuclear spin polarization back-acts on the
electron spin via hyperfine interaction, described by the
Hamiltonian $\hat{H}_{hf}=A(\hat{s} \cdot \hat{I})$. The hyperfine
constant $A$ is positive for Ga and As nuclei due to their
positive gyromagnetic ratios $\gamma>0$. As a result the
$\Uparrow\downarrow$ exciton shifts to higher energy under
$\sigma^+$ excitation. In a similar manner, under $\sigma^-$
excitation [Fig.~\ref{Fig:FigSSign}(c)] the population of the
$\Downarrow\uparrow$ exciton is enhanced, and it also shifts to
higher energy since now both $s_z$ and $I_z$ are positive. This is
indeed observed in Fig.~\ref{Fig:FigSSign}(a): the Zeeman
component whose intensity is enhanced by circularly polarized
excitation always shifts to higher energy. This observation
confirms the positive sign of $A$ and that the spin flip-flops are
the source of dynamic nuclear spin polarization. Taking also into
account the signs of the electron and hole $g$-factors, we
conclude that circularly polarized excitation that enhances the
high- (low-) energy exciton population and labeled here
$\sigma^+$($\sigma^-$), populates predominantly nuclear spin
states with negative (positive) projection $I_z$.

For the NMR spectra measured with $\sigma^+$ optical excitation
the population of the $I_z=-3/2$ and $I_z=-1/2$ states is enhanced
as discussed above. As a result the amplitude of the
$-3/2\leftrightarrow-1/2$ satellite NMR peak exceeds the amplitude
of the $+1/2\leftrightarrow+3/2$ satellite \cite{STsAsymmetry}.
The spectra of Fig.~\ref{Fig:FigRawData}(b) were measured under
$\sigma^+$ optical excitation (i.e., excitation that enhances the
intensity of the high energy Zeeman exciton component). For the
case of compressive stress along $[001]$ the
$-3/2\leftrightarrow-1/2$ NMR peak has a lower frequency than the
$-1/2\leftrightarrow+1/2$ central peak [middle spectrum in
Fig.~\ref{Fig:FigRawData}(b)], corresponding to $\nu_Q<0$. The
quadrupolar shift $\nu_Q$ is related to strain via
Eqs.~\ref{Eq:nuNMR32}, \ref{Eq:nuQAppendix} and we find that
$\nu_Q=\frac{eQ}{2 h}S_{11}\epsilon_{b}<0$ in this experiment. In
case of $[001]$ compression $\epsilon_{b}<0$, and since the
quadrupolar moment of $^{75}$As is positive\cite{STONE20161}
$Q>0$, we concluded that $S_{11}>0$ for $^{75}$As.

While in the above calculations we assumed $B_z>0$, the opposite
assumption $B_z<0$ leads to the same conclusions about the signs
of the gradient elastic tensor components $S_{11}$. Finally, we
note that in previous work on InGaAs/GaAs\cite{chekhovich2012} and
GaAs/AlGaAs\cite{PhysRevB.93.165306} QDs, the shift of the
$-3/2\leftrightarrow-1/2$ satellite peak to lower frequency was
arbitrarily assigned a positive $\nu_Q$ value since the sign of
$S_{11}$ was undefined. By contrast, in the present work the sign
of $\nu_Q$ is strictly determined by
Eqs.~\ref{Eq:nuQ},~\ref{Eq:nuNMR32}, \ref{Eq:nuQAppendix} and the
signs of $S_{11}$ and $Q$.

\subsection{Measurement of the ratio of the electric field
gradients on $As$ and $Ga$ lattice sites\label{SubSec:GavsAs}}

\begin{figure}[t]
\includegraphics[width=1.0\linewidth]{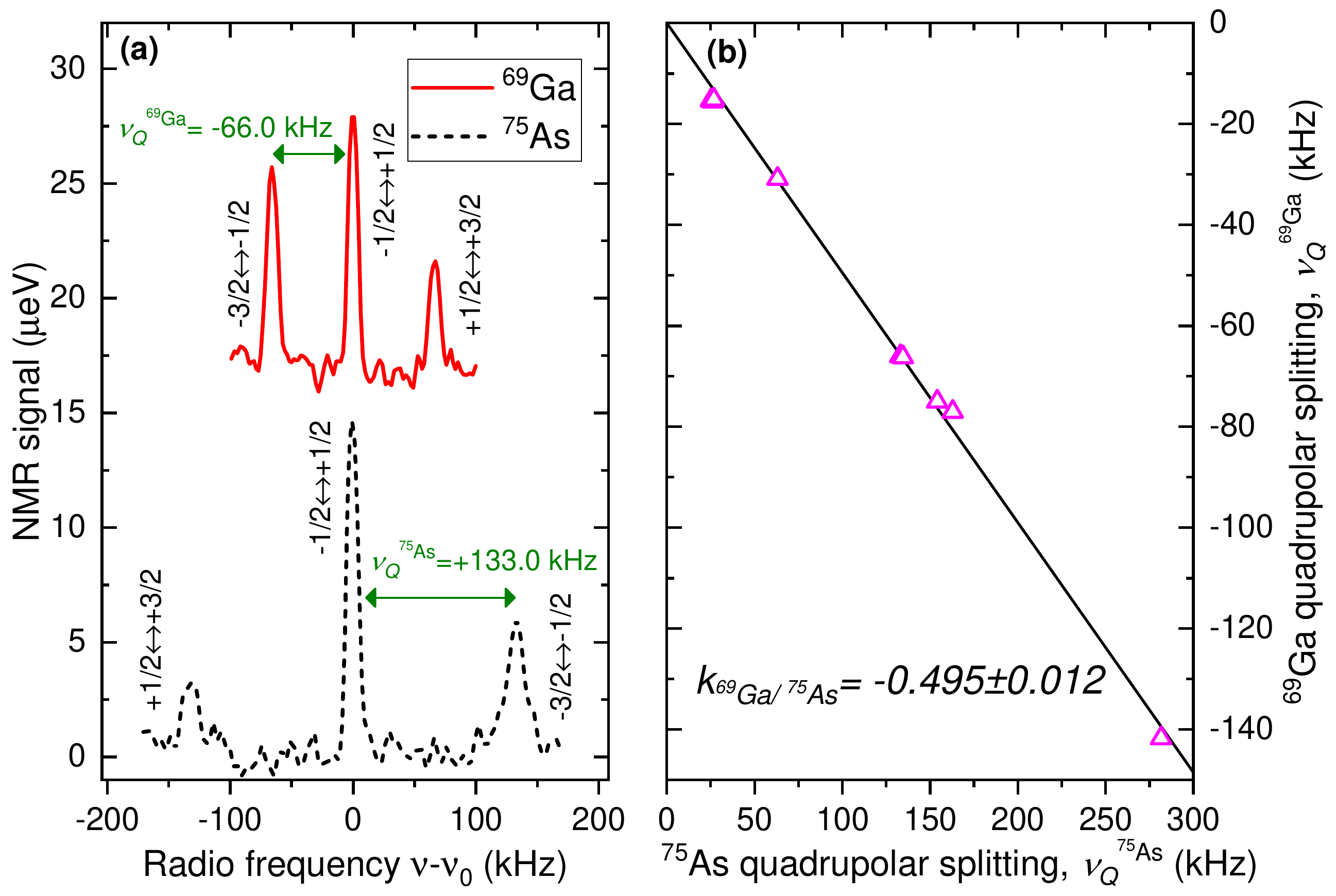}
\caption{\label{Fig:FigGavsAs} (a) NMR spectra of a single quantum
dot measured at high magnetic field $B_z\approx5.5$~T on $^{75}$As
nuclei ($\nu_0\approx$40.27~MHz, dashed line) and $^{69}$Ga nuclei
($\nu_0\approx$56.46~MHz, solid line). Both isotopes are spin-3/2
and exhibit well resolved quadrupolar triplets with different
splittings $\nu_Q$. (b) Dependence of the $^{69}$Ga quadrupolar
splitting $\nu_Q^{^{69}\mathrm{Ga}}$ on the $^{75}$As splitting
$\nu_Q^{^{75}\mathrm{As}}$ measured on different individual
quantum dots in an unstressed sample as well as in samples
stressed along $[001]$ or $[110]$ crystal axes (symbols). Linear
fitting is shown by the solid line and its slope
$k_{^{69}\mathrm{Ga}/^{75}\mathrm{As}}=-0.495\pm0.012$ gives an
estimate of the ratio
$(Q^{^{69}\mathrm{Ga}}S_{11}^{^{69}\mathrm{Ga}})/(Q^{^{75}\mathrm{As}}S_{11}^{^{75}\mathrm{As}})$
between the products of gradient elastic tensor components
$S_{11}$ and quadrupolar moments $Q$ of $^{69}$Ga and $^{75}$As in
GaAs.}
\end{figure}

Measurement of NMR via optical detection of the hyperfine shifts
in the PL spectra of a quantum dot guarantees that only the nuclei
of a single quantum dot contribute to the NMR spectrum. Thus if
NMR spectrum is measured on As and Ga nuclei of the same quantum
dot, one can ensure that the nuclei of the two isotopes belong to
the same nanoscale volume and probe the same strain field. Then,
according to Eq.~\ref{Eq:nuQ}, the ratio of the quadrupolar shifts
of the two isotopes is simply
$\nu_Q^{^{69}\mathrm{Ga}}/\nu_Q^{^{75}\mathrm{As}}=(Q^{^{69}\mathrm{Ga}}S_{11}^{^{69}\mathrm{Ga}})/(Q^{^{75}\mathrm{As}}S_{11}^{^{75}\mathrm{As}})$
and does not depend on the actual strain magnitude $\epsilon_b$.
Figure~\ref{Fig:FigGavsAs}(a) shows NMR spectra of $^{69}$Ga (top)
and $^{75}$As (bottom) measured on the same quantum dot at
$B_z=$5.5~T using $\sigma^+$ optical excitation. Both isotopes
have spin $I=$3/2 giving rise to the well resolved NMR triplets
with different quadrupolar splittings $\nu_Q$. We note that the
satellite peak with higher amplitude, corresponding to the
$-3/2\leftrightarrow-1/2$ transition, appears on the low (high)
frequency side for Ga (As) implying opposite signs of $\nu_Q$ and
hence opposite signs of $QS_{11}$ for the two isotopes. Since
$S_{11}>0$ for $^{75}$As and $Q>0$ for all stable Ga and As
isotopes we conclude that $S_{11}<0$ for $^{69}$Ga and $^{71}$Ga.
Similar measurements of $^{69}$Ga and $^{75}$As NMR were conducted
on several quantum dots in an unstressed and stressed samples and
are summarized in Fig.~\ref{Fig:FigGavsAs}(b) where
$\nu_Q^{^{69}\mathrm{Ga}}$ is shown as a function of
$\nu_Q^{^{75}\mathrm{As}}$ by the symbols. The linear fit is shown
by the line and yields the slope
$k_{^{69}\mathrm{Ga}/^{75}\mathrm{As}}=(Q^{^{69}\mathrm{Ga}}S_{11}^{^{69}\mathrm{Ga}})/(Q^{^{75}\mathrm{As}}S_{11}^{^{75}\mathrm{As}})=-0.495\pm0.012$.
Taking the values of quadrupolar moments \cite{STONE20161}
$Q^{^{69}\mathrm{Ga}}=0.171\times10^{-28}$~m$^2$ and
$Q^{^{75}\mathrm{As}}=0.314\times10^{-28}$~m$^2$ we calculate for
the ratio of the components of the gradient elastic tensors:
$S_{11}^{^{69}\mathrm{Ga}}/S_{11}^{^{75}\mathrm{As}}=-0.909$ so
that the magnitude of the strain-induced EFG is smaller at the
gallium sites.

\subsection{Derivation of the gradient-elastic tensor component $S_{11}$ in GaAs\label{Subsec:SDerivation}}

We now discuss how simultaneous measurements of GaAs free exciton
PL and QD NMR presented in Section~\ref{SubSec:RawData} can be
used to calibrate the fundamental material parameters of GaAs.
First we consider the case of a sample stressed along the $[001]$
direction. If the stress is produced by applying a uniform
$z$-oriented pressure to the top and bottom $(001)$ surfaces of
the sample, the resulting strain has a very simple configuration
where $\epsilon_h$ and $\epsilon_b$ are finite, while
$\epsilon_\eta$ and $\epsilon_s$ vanish for symmetry reasons
($\epsilon_{xx}=\epsilon_{yy}$ and
$\epsilon_{xy}=\epsilon_{yz}=\epsilon_{zx}=0$). In this case
according to Eq.~\ref{Eq:EPL} the splitting between the LH and HH
PL transition energies is simply $2|b\epsilon_b|$. Since both the
LH-HH exciton splitting and the NMR shift now depend only on
$\epsilon_b$, their ratio can be taken to eliminate $\epsilon_b$
and we find:
\begin{equation}
\begin{split}
k_{[001]}^-=-\frac{4hb}{eQS_{11}},
\end{split}
\label{Eq:k001-}
\end{equation}
where the minus sign is added to account for the fact that the PL
of the predominantly LH exciton has a lower energy at
$\epsilon_b<0$. The $k_{[001]}^-$ has been measured experimentally
for $^{75}$As (see Fig.~\ref{Fig:FigEPLMeanDiff}(a) and
Section~\ref{SubSec:RawData}).

In a real sample, the pressure on the surfaces of the sample is
not necessarily uniform and aligned to the $z$ axis. In this case
Eq.~\ref{Eq:k001-} holds only at the geometrical centre of the top
$(001)$ surface (for symmetry reasons), while away from the centre
the non-diagonal strain components may arise leading e.g. to
$\epsilon_s\ne0$. According to Eqs.~\ref{Eq:EPL}, \ref{Eq:nuQ} the
effect of the finite $\epsilon_s$ or $\epsilon_\eta$ is to induce
an additional splitting of the heavy and light hole excitons
without affecting the NMR spectral splitting $\nu_Q$, in which
case the dependence of the LH-HH splitting on $\nu_Q$ would no
longer be linear when measured over the surface of the sample. In
experiment, multiple spots of the $[001]$-stressed sample, both at
the centre of the sample surface and close to the edges were
investigated. The strain can be seen to vary significantly across
the sample surface: $\nu_Q$ is found to range between
$-$400..$-$180~kHz [Fig.~\ref{Fig:FigEPLMeanDiff}(a)] indicating
variation of $\epsilon_b$, while the spread in the average GaAs PL
energies [full triangles in Fig.~\ref{Fig:FigEPLMeanDiff}(a)]
indicates variation of $\epsilon_h$. On the other hand the
resulting dependence of the LH-HH splitting on $\nu_Q$ is still
well described by a linear function [open triangles and dashed
line in Fig.~\ref{Fig:FigEPLMeanDiff}(a)]. This can only be if
$\epsilon_s$ and $\epsilon_\eta$ are small in the studied sample
and thus the experimentally measured ratio $k_{[001]}^-=46.5\pm0.8
\mu$eV/kHz describes the relation of the fundamental parameters
$b$ and $S_{11}$ of GaAs according to Eq.~\ref{Eq:k001-}.

We now consider the case of a sample stressed along the $[110]$
direction. If the stress is produced by applying a uniform
pressure along $[110]$ to the $(110)$ surfaces of the sample the
resulting strain will have non-zero $\epsilon_h$, $\epsilon_b$, as
well as $\epsilon_s$ arising from the $\epsilon_{xy}$ component.
(Recall that $x$ and $y$ axes are aligned along $[100]$ and
$[010]$ respectively, so that
$\epsilon_{xy}=\epsilon_{xx}=\epsilon_{yy}$ under uniform stress
along $[110]$). Even in an ideal case the GaAs PL energies
(Eq.~\ref{Eq:EPL}) under $[110]$ stress involve the
$d^2\epsilon_s^2$ term, making it difficult to relate to the NMR
shifts given by Eq.~\ref{Eq:nuQ}. In a real sample, the
non-diagonal shear strain $\epsilon_{xy}$ is not necessarily
constant and $\epsilon_\eta$ is not necessarily zero due to the
inevitable non-uniformities of the stress induced by the titanium
strain mount. This is evidenced in
Fig.~\ref{Fig:FigEPLMeanDiff}(b) where the LH-HH splitting (open
squares) is seen to deviate considerably from a linear dependence
on $\nu_Q$, which is proportional only to $\epsilon_b$.

However, it is possible to eliminate the effect of the unknown
shear strain components $\epsilon_s$, $\epsilon_\eta$ in a
$[110]$-stress configuration. For that we notice that the top
$(001)$ surface of the sample which is studied optically is free
from external stress (traction free). As a result, the boundary
conditions dictate\cite{BarberBook} that three of the components
of the mechanical stress tensor vanish
$\sigma_{zx}=\sigma_{zy}=\sigma_{zz}=0$, and the only non-zero
components are $\sigma_{xx}$, $\sigma_{yy}$, $\sigma_{xy}$.
Writing down the strain-stress relation one can easily verify that
in a GaAs crystal (cubic symmetry), the ratio of the biaxial and
hydrostatic strains at the free $(001)$ surface does not depend on
the actual $\sigma_{xx}$, $\sigma_{yy}$, $\sigma_{xy}$ values and
equals
$\epsilon_b/\epsilon_h=\frac{\sigma_{xx}+\sigma_{yy}}{2c_{12}+c_{11}}/\frac{\sigma_{xx}+\sigma_{yy}}{2c_{12}-2c_{11}}=\frac{2c_{12}+c_{11}}{2c_{12}-2c_{11}}\approx-1.742$,
where $c_{11}$ and $c_{12}$ are the stiffness constants of
GaAs\cite{Adachi2009}. Now we use this relation to express
$\epsilon_b$ through $\epsilon_h$ in Eq.~\ref{Eq:nuQ} to make
quadrupolar shift $\nu_Q$ depend only on $\epsilon_h$. Since the
average LH-HH shift of the GaAs PL energy $a\epsilon_h$ also
depends only on $\epsilon_h$ (Eq.~\ref{Eq:EPL}) it can be related
to $\nu_Q$ by eliminating the strain to find:
\begin{equation}
\begin{split}
k_{[110]}^+=\frac{4ha}{eQS_{11}}\frac{c_{12}-c_{11}}{2c_{12}+c_{11}},
\end{split}
\label{Eq:k110+}
\end{equation}
The solid symbols and the line in Fig.~\ref{Fig:FigEPLMeanDiff}(b)
demonstrate that the average GaAs PL energy is indeed a linear
function of $\nu_Q$, confirming the invariance of
$\epsilon_b/\epsilon_h$ at the surface of the studied sample. Thus
Eq.~\ref{Eq:k110+} relates the $a$ and $S_{11}$ parameters through
the experimentally measured value
$k_{[110]}^+=55.1\pm1.5~\mu$eV/kHz.

\begin{table*}[t]
\caption{GaAs parameters.\label{Tab:GaParams2}}
\begin{ruledtabular}
\begin{tabular}{l|l|c|c}
Parameter & Units & Previous work & This work \\
\hline
$a=a_c+a_v$ & eV &$-8.7$~\cite{PhysRevLett.16.942}; $-8.9$~\cite{PhysRev.161.695}; $-8.7$~\cite{BENDORIUS19701111}; $-8.93$~\cite{PhysRevB.15.2127}; $-8.72$~\cite{QIANG19901087}; $-10.19$~\cite{MAIR1998277}; $-8.5$~\cite{Vurgaftman2001}; $-8.8$~\cite{Adachi2009}&\\
\hline
$b$ & eV & $-2.1$~\cite{PhysRevLett.16.942}; $-1.96$~\cite{PhysRev.161.695}; $-1.76$~\cite{PhysRevB.15.2127}; $-2.00$~\cite{QIANG19901087}; $-2.00$~\cite{MAIR1998277}; $-2.0$~\cite{Vurgaftman2001}; $-1.85$~\cite{Adachi2009}&\\
\hline
$b/a$ & & $0.24$~\cite{PhysRevLett.16.942}; $0.22$~\cite{PhysRev.161.695}; $0.20$~\cite{PhysRevB.15.2127}; $0.23$~\cite{QIANG19901087}; $0.20$~\cite{MAIR1998277}; $0.24$~\cite{Vurgaftman2001}; $0.21$~\cite{Adachi2009}& ${0.241\pm0.008}$\\
\hline
$c_{11}$ & GPa & $122.1$~\cite{Vurgaftman2001}; $118.8$~\cite{Adachi2009}&\\
\hline
$c_{12}$ & GPa &$56.6$~\cite{Vurgaftman2001}; $53.8$~\cite{Adachi2009}&\\
\hline
$c_{44}$ & GPa &$60.0$~\cite{Vurgaftman2001}; $59.4$~\cite{Adachi2009}&\\
\hline
$\frac{2c_{12}+c_{11}}{2c_{12}-2c_{11}}$ & &$-1.796$~\cite{Vurgaftman2001}; $-1.742$~\cite{Adachi2009}&\\
\hline
$\frac{Q^{^{69}\mathrm{Ga}}S_{11}^{^{69}\mathrm{Ga}}}{Q^{^{75}\mathrm{As}}S_{11}^{^{75}\mathrm{As}}}$
& & $-0.508$~\cite{PhysRevB.10.4244}&${-0.495\pm0.012}$\\
\hline $Q^{^{75}\mathrm{As}}S_{11}^{^{75}\mathrm{As}}$ &
$10^{-6}$~V& $\pm1.06$~\cite{PhysRevB.10.4244}&${+0.76}$\\
\hline $Q^{^{69}\mathrm{Ga}}S_{11}^{^{69}\mathrm{Ga}}$ &
$10^{-6}$~V& $\mp0.543$~\cite{PhysRevB.10.4244}&${-0.37}$\\
\hline
$Q^{^{69}\mathrm{Ga}}$ & $10^{-28}$~m$^2$ &$0.171$~\cite{STONE20161}&\\
\hline
$Q^{^{75}\mathrm{As}}$ & $10^{-28}$~m$^2$ &$0.314$~\cite{STONE20161}&\\
\hline
$S_{11}^{^{75}\mathrm{As}}$ & $10^{21}$~V/m$^2$ & $\pm34.0$~\cite{PhysRevB.10.4244}&${+24.1}$\\
\hline
$S_{11}^{^{69}\mathrm{Ga}}$ & $10^{21}$~V/m$^2$ & $\mp31.7$~\cite{PhysRevB.10.4244}&${-21.9}$\\
\end{tabular}
\end{ruledtabular}
\end{table*}

Since the absolute values of stress and strain are not measured in
our experiment, the results presented above can be used to
estimate the ratios of the GaAs parameters. The absolute value of
a parameter can then be estimated by taking the values of other
parameters from the previous studies.

The elementary charge $e$ and the Planck constant $h$ are known
with very high accuracy. The stiffness constants of GaAs $c_{11}$
and $c_{12}$ are also known with a good accuracy. While $c_{11}$
and $c_{12}$ in GaAs exhibit some temperature dependence, the
ratio $c_{11}/c_{12}$ used in our analysis is reported to be
nearly invariant from cryogenic to room temperature
\cite{Cottam1973}. Here we use the $c_{11}$, $c_{12}$ values at
300~K from Ref.~\cite{Adachi2009}. This leaves three GaAs
parameters: the deformation potentials $a=(a_c+a_v)$, $b$ and the
$QS_{11}$ product of $^{75}$As. Since there are two experimentally
measured ratios (Eqs.~\ref{Eq:k001-}, \ref{Eq:k110+}), these three
parameters can be linked by two independent relations.

One of the relations can be obtained by dividing
Eqs.~\ref{Eq:k001-} and \ref{Eq:k110+} to eliminate $QS_{11}$
which gives the ratio of the deformation potentials:
\begin{equation}
\begin{split}
&\frac{b}{a}=\frac{k_{[100]}^-}{k_{[110]}^+}\frac{c_{12}-c_{11}}{2c_{12}+c_{11}}=\\
&\quad\quad=(0.841\pm0.027)\frac{c_{12}-c_{11}}{2c_{12}+c_{11}}=0.241\pm0.008,
\end{split}
\label{Eq:bVSa}
\end{equation}
where the error estimate is purely due to the experimental
uncertainty in $k_{[100]}^-$ and $k_{[110]}^+$ and there can be an
additional error due to the $\sim\pm$2\% uncertainty in the
$c_{11}$, $c_{12}$ values. Our estimate of $b/a$ is in excellent
agreement with the ratio derived from the recommended
\cite{Vurgaftman2001,Adachi2009} values of $a$ and $b$ based on a
number of independent experimental and theoretical studies. Such
agreement supports the validity of our experimental method based
on relating PL and NMR spectral shifts. The estimates derived in
this work are summarized in Table~\ref{Tab:GaParams2} together
with the results of the earlier work.

For the second relation we use Eq.~\ref{Eq:k110+} to link the
deformation potential $a$ with the component of the gradient
elastic tensor $S_{11}$. The variation of the GaAs fundamental gap
under hydrostatic strain characterized by $a=a_c+a_v$ has been
studied experimentally by several
authors\cite{PhysRevLett.104.067405,doi:10.1063/1.1566463,doi:10.1063/1.2204843,doi:10.1002/adma.201200537,doi:10.1002/pssb.201100775}.
There is some variation, but most experiments as well as
calculations \cite{PhysRevB.39.1871} are consistent and it is
commonly accepted\cite{Vurgaftman2001,Adachi2009} that
$a\approx-8.8$~eV. By contrast there are only few reports on
experimental gradient-elastic tensors in GaAs
\cite{PhysRevB.10.4244,Bogdanov1968,Guerrier1997} with only one
series of experiments where $S_{11}$ and $S_{44}$ were measured
directly \cite{PhysRev.177.1221,PhysRevB.10.4244}. Thus we use
$a\approx-8.8$~eV (Ref.\cite{Adachi2009}) to evaluate
$Q^{^{75}\mathrm{As}}S_{11}^{^{75}\mathrm{As}}\approx+0.76$~$\mu$V.
The uncertainty of this estimate arising from the experimental
uncertainty in $k_{[110]}^+$ is only $\pm3\%$, so the main error
is likely to arise from the uncertainty in $a$, which is
approximately $\pm5\%$ based on the spread of the values derived
in different independent studies. Using the
$k_{^{69}\mathrm{Ga}/^{75}\mathrm{As}}$ ratio measured in
Sec.~\ref{SubSec:GavsAs} we also estimate
$Q^{^{69}\mathrm{Ga}}S_{11}^{^{69}\mathrm{Ga}}\approx-0.37$~$\mu$V
for $^{69}$Ga with a similar relative uncertainty. These values
are $\sim$30\% smaller than those derived by
Sundfors\cite{PhysRevB.10.4244} from the nuclear acoustic
resonance measurements (Table~\ref{Tab:GaParams2}). We point out
here that normally it is the $QS_{11}$ product and not $S_{11}$
that is measured in NMR experiments and is used to predict the NMR
spectra in strained semiconductor structures -- the individual
values for $S_{ijkl}$ and $Q$ are not accessible in conventional
NMR measurements. Nonetheless, for the reference, we quote in
Table~\ref{Tab:GaParams2} the $S_{11}$ values derived in this work
and reported in Ref.\cite{PhysRevB.10.4244}, where in both case we
divided the measured $QS_{11}$ products by the most recent
recommended values\cite{STONE20161} of quadrupolar moments $Q$.
For practical applications, it is preferable to use the $QS_{11}$
product values, or when using $S_{ijkl}$ and $Q$ separately, take
their values from the same source. We also note that the $S_{11}$
values in Ref.\cite{PhysRevB.10.4244} are in the c.g.s. units of
$\times10^{15}$ statcoulomb/cm$^3$ and are multiplied here by
2997924.580 to convert them to the V/m$^2$ SI units.

\section{Discussion and conclusion}

An important feature of this work is that elastic strain is probed
optically through the spectral shifts the free exciton PL in bulk
GaAs. This method offers certain advantages: there is no need to
measure the stress or control precisely the size and the shape of
the sample, moreover the strain can be probed locally on a
micrometer scale, so that modest strain inhomogeneities across the
sample are not a limitation. The downside is that the accuracy of
the measured strain is limited by the current uncertainty in the
deformation potentials. On the other hand, the detection of the
nuclear quadrupolar effects in this work is achieved in a most
straightforward way -- by measuring the quadrupolar splitting of
the NMR spectral triplet. This is different from the previous
studies on GaAs \cite{PhysRev.177.1221,PhysRevB.10.4244} where
detection was rather indirect and relied on measuring the changes
of the quality factors of mechanical resonances.

The $\sim30$\% difference in the measured $S_{11}$ values between
this work and the work of Sundfors
\cite{PhysRev.177.1221,PhysRevB.10.4244} appears to be too large
to be attributed to the uncertainty in deformation potentials of
GaAs. On the other hand we note that the ratio of $S_{11}$ for Ga
and As is in remarkably good agreement. Moreover, it was pointed
out by Sundfors \cite{PhysRev.177.1221} that his room temperature
acoustic resonance measurements of $S_{11}$ for $^{115}$In in InSb
were notably larger than the corresponding $S_{11}$ values
obtained in two independent studies using static strain
\cite{PhysRev.107.953,Bogdanov1968} at 77~K. One possibility, is
that all of the $S_{ijkl}$ values reported in
Refs.\cite{PhysRev.177.1221,PhysRevB.10.4244} had a systematic
offset arising from a number of parameters that needed to be
calibrated for acoustic resonance measurements. Moreover, the
deviation in the results may arise from the fundamental
differences in how nuclear spin system responds to static and
dynamic (acoustic wave) strain, as well as from the temperature
dependence -- these aspects remain unexplored and would require
further work.

The PL/NMR method for derivation of the gradient-elastic tensor
reported here have potential to be extended further. For example
the $S_{44}$ component of GaAs that was not probed here, can be
measured. Such a measurement would require shear strain and
magnetic field which is not parallel to one of the cubic axes
(e.g. $[001]$). The GaAs/AlGaAs pair is unique since it permits
nearly lattice matched epitaxial growth. As a result external
stress can induce deformations significantly exceeding the
built-in strain, making it possible to use bandgap shifts to gauge
the strain. Application to other materials, e.g. InAs/GaAs quantum
wells and dots may require alternative methods for probing the
strain, such as X-ray diffraction.

For practical applications the $QS_{11}$ for GaAs can be taken
directly from the values measured here
(Table~\ref{Tab:GaParams2}). For the $QS_{44}$ parameters that
were not measured here, we recommend taking the values from
Ref.\cite{PhysRevB.10.4244} and rescaling by a factor of $0.7$,
which is the ratio of the $QS_{11}$ values measured here and in
Ref.\cite{PhysRevB.10.4244}. Since GaAs and InAs were found to
have very similar gradient elastic tensors\cite{PhysRevB.10.4244},
similar scaling by a factor of $0.7$ can be applied to the
$S$-tensor values for InAs.

\section*{Acknowledgements}
The authors are grateful to Ceyhun Bulutay and Yongheng Huo for
fruitful discussions. This work was supported by the EPSRC
Programme Grant EP/N031776/1, the Linz Institute of Technology
(LIT) and the Austrian Science Fund (FWF): P 29603. E.A.C. was
supported by a Royal Society University Research Fellowship.

\appendix

\section{Relation between strain and nuclear quadrupolar effects\label{Appendix:QNuc}}

The second spatial derivatives of the electrostatic potential
$V(x,y,z)$ at the nuclear sites 
form a second rank symmetric tensor
$V_{\alpha\beta}=\frac{\partial^2 V}{\partial\alpha\partial\beta}$
($\alpha,\beta=x,y,z$). Small deformation of a solid body is
described via the second rank elastic strain tensor
\begin{equation}
\begin{split}
\epsilon_{ij}=\frac{\partial u_i}{\partial x_j} \quad(i,j=x,y,z),
\end{split}
\label{Eq:epsilon}
\end{equation}
where $u_i$ are the components of the vector field of
displacements $\vec{u}(x,y,z)$ characterizing the deformation. In
the limit of small deformation $V_{ij}$ is related to
$\epsilon_{kl}$ via:
\begin{equation}
\begin{split}
V_{ij}=\sum_{k,l}S_{ijkl}\epsilon_{kl} \quad(i,j,k,l=x,y,z),
\end{split}
\label{Eq:V=Se}
\end{equation}
where $S_{ijkl}$ is a fourth rank ''gradient-elastic'' tensor. Not
all of its 81 components are independent, and the number of
independent parameters is greatly reduced further in crystal
structures with high symmetry. In case of a zinc-blend crystal
(cubic symmetry group $T_d$) the non-vanishing elements of
$S_{ijkl}$ are \cite{PhysRev.107.953}:
\begin{equation}
\begin{split}
&S_{xxxx}=S_{yyyy}=S_{zzzz}\\
&S_{yzyz}=S_{zxzx}=S_{xyxy}\\
&S_{xxyy}=S_{yyzz}=S_{zzxx}=S_{xxzz}=S_{zzyy}=S_{yyxx}\\
\end{split}
\end{equation}
Moreover, since $V_{ij}$ and $\epsilon_{ij}$ are both symmetric,
the gradient elastic tensor has an additional symmetry with
respect to the pari of the first and second indices as well as the
pair of the third and fourth indices
($S_{ijkl}=S_{jikl}=S_{ijlk}=S_{jilk}$). Thus in a coordinate
frame aligned with the crystal axes $x\parallel[100]$,
$y\parallel[010]$, $z\parallel[001]$ there are in total 21
non-zero components and the tensor is fully characterized by 3
independent parameters $S_{xxxx}$, $S_{xxyy}$ and $S_{yzyz}$.
Taking into account the symmetries of $S_{ijkl}$ we can evaluate
Eq.~\ref{Eq:V=Se} to find the explicit expression for the electric
field gradients:
\begin{equation}
\begin{split}
&V_{1}=V_{xx}=S_{xxxx}\epsilon_{xx}+S_{xxyy}(\epsilon_{yy}+\epsilon_{zz})=\\
&\quad=S_{xxxx}(\epsilon_{xx}-(\epsilon_{yy}+\epsilon_{zz})/2)\\
&V_{2}=V_{yy}=S_{xxxx}\epsilon_{yy}+S_{xxyy}(\epsilon_{xx}+\epsilon_{zz})=\\
&\quad=S_{xxxx}(\epsilon_{yy}-(\epsilon_{xx}+\epsilon_{zz})/2)\\
&V_{3}=V_{zz}=S_{xxxx}\epsilon_{zz}+S_{xxyy}(\epsilon_{xx}+\epsilon_{yy})=\\
&\quad=S_{xxxx}(\epsilon_{zz}-(\epsilon_{xx}+\epsilon_{yy})/2)\\
&V_{4}=V_{yz}=V_{zy}=2S_{yzyz}\epsilon_{yz}\\
&V_{5}=V_{xz}=V_{zx}=2S_{yzyz}\epsilon_{xz}\\
&V_{6}=V_{xy}=V_{yx}=2S_{yzyz}\epsilon_{xy}
\end{split}
\label{Eq:VCubic}
\end{equation}
where the right hand side parts of the first three equations were
obtained by setting $S_{xxyy}=-S_{xxxx}/2$, which is a common
convention to take into account the fact that only the traceless
part of $V_{ij}$ is observable in NMR \cite{PhysRev.128.1630}. We
have also introduced EFG components $V_{m}\quad(m=1..6)$ in Voigt
notation, using which we can rewrite Eq.~\ref{Eq:VCubic} as:
\begin{equation}
\begin{split}
&V_{1}=S_{11}(\epsilon_{1}-(\epsilon_{2}+\epsilon_{3})/2)\\
&V_{2}=S_{11}(\epsilon_{2}-(\epsilon_{1}+\epsilon_{3})/2)\\
&V_{3}=S_{11}(\epsilon_{3}-(\epsilon_{1}+\epsilon_{2})/2)\\
&V_{4}=S_{44}\epsilon_{4}\\
&V_{5}=S_{44}\epsilon_{5}\\
&V_{6}=S_{44}\epsilon_{6},
\end{split}
\label{Eq:VCubicVoigt}
\end{equation}
where $S_{11}=S_{xxxx}$ and $S_{44}=S_{yzyz}$. While Voigt
notation simplifies the equations and is commonly accepted it
needs to be used with care. Unlike $S_{ijkl}$, the 2$\times$2
matrix $S_{mn}$ is not a tensor and does not follow the tensor
transformation rules. One of the consequences of this is that the
definition of the non-diagonal components of strain should include
an additional factor of $2$, so that
$\epsilon_{4}=2\epsilon_{yz}$, $\epsilon_{5}=2\epsilon_{xz}$,
$\epsilon_{6}=2\epsilon_{xy}$, while this factor of 2 is not
needed in the definition of $V_{4}$, $V_{5}$, $V_{6}$ (see
Eq.~\ref{Eq:VCubic}). A similar situation is encountered in the
strain-stress relation $\sigma_{ij}=c_{ijkl}\epsilon_{kl}$
expressed in Voigt notation where the shear strains
$\epsilon_{4}$, $\epsilon_{5}$, $\epsilon_{6}$ require a factor of
2 in their definition, while there is no such factor for the
stress components $\sigma_{4}$, $\sigma_{5}$, $\sigma_{6}$ (see
Ch.~10 in Ref.~\cite{NewnhamBook}).

The Hamiltonian $\hat{H}_{Q}$ describing the interaction of the
nucleus with spin $I$ and quadrupolar moment Q with the electric
field gradients is\cite{SlichterBook}:
\begin{equation}
\begin{split}
\hat{H}_{Q}=\frac{eQ}{6I(2I-1)h}\sum_{i,j=x,y,z}V_{ij}\left(\frac{3}{2}(\hat{I}_i\hat{I}_j+\hat{I}_j\hat{I}_i)-\delta_{ij}I^2\right),
\end{split}
\label{Eq:HQ}
\end{equation}
where $e>0$ is the elementary charge, $h$ is the Planck constant,
$\delta_{ij}$ is Kronecker's delta, $\hat{I_i}$ are spin operator
components in Cartesian coordinates and the Hamiltonian is in
frequency units (Hz). Static magnetic field gives rise to the
Zeeman Hamiltonian
\begin{equation}
\begin{split}
\hat{H}_{Z}=-\frac{\gamma B_{z}}{2\pi}\hat{I}_z,
\end{split}
\label{Eq:HZ}
\end{equation}
where $\gamma$ is the nuclear gyromagnetic ratio, and we
explicitly consider the case of the field $B_z$ aligned along the
$z$ axis. For the spin-3/2 nuclei the total Hamiltonian $H_Z+H_Q$
is a 4$\times$4 matrix and can in principle be diagonalised
analytically to find the eigenstates.

A much simpler approximate solution can be found for the case of
large magnetic field. In our experiments the effects induced by
magnetic field (characterized by Larmor frequency $>$40~MHz) are
at least two orders of magnitude larger than the quadrupolar
effects (characterized by quadrupolar shifts $<$0.4~MHz). Thus
with good accuracy quadrupolar effects can be treated as a
perturbation, and to the first order we can omit all off-diagonal
terms of the total Hamiltonian\cite{PhysRev.79.685}. The resulting
eigenstates are the eigenstates of the $\hat{I}_z$ operator with
eigenenergies (in Hz units):
\begin{equation}
\begin{split}
&E_{-3/2}=\frac{3\gamma B_{z}}{4\pi}+\frac{eQ}{4h}S_{11}\epsilon_{b}\\
&E_{-1/2}=\frac{\gamma B_{z}}{4\pi}-\frac{eQ}{4h}S_{11}\epsilon_{b}\\
&E_{+1/2}=-\frac{\gamma B_{z}}{4\pi}-\frac{eQ}{4h}S_{11}\epsilon_{b}\\
&E_{+3/2}=-\frac{3\gamma
B_{z}}{4\pi}+\frac{eQ}{4h}S_{11}\epsilon_{b},
\end{split}
\label{Eq:En32}
\end{equation}
where we have substituted the EFG values from Eq.~\ref{Eq:VCubic},
the energies are indexed by their corresponding $\hat{I}_z$
eigenvalue, and the effect of elastic deformation on the nuclear
spin states is manifested only via the ''biaxial'' part of strain
$\epsilon_{b}=\epsilon_{zz}-(\epsilon_{xx}+\epsilon_{yy})/2$. The
dipolar transitions are allowed for the pairs of states where
$I_z$ changes by $\pm$1, and the NMR frequencies are obtained by
taking the differences of the corresponding energies in
Eq.~\ref{Eq:En32}:
\begin{equation}
\begin{split}
&\nu _{-3/2\leftrightarrow -1/2}=\frac{\gamma B_z}{2 \pi}+\frac{eQ}{2 h}S_{11}\epsilon_{b}\\
&\nu _{-1/2\leftrightarrow +1/2}=\frac{\gamma B_z}{2 \pi}\\
&\nu _{+1/2\leftrightarrow +3/2}=\frac{\gamma
B_z}{2\pi}-\frac{eQ}{2 h}S_{11}\epsilon_{b}.\\
\end{split}
\label{Eq:nuNMR32}
\end{equation}
Equation~\ref{Eq:nuNMR32} describes a triplet of NMR transitions
with a central transition $-1/2\leftrightarrow +1/2$ unaffected by
strain and two satellite transitions on either side of the central
transition, separated by the quadrupolar shift
\begin{equation}
\begin{split}
\nu_{Q}=\frac{eQ}{2 h}S_{11}\epsilon_{b},\\
\end{split}
\label{Eq:nuQAppendix}
\end{equation}
which is the same as Eq.~\ref{Eq:nuQ}.


%

\end{document}